# Wikipedia Page View Reflects Web Search Trend

Mitsuo Yoshida
yoshida@cs.tut.ac.jp
Toyohashi
University of Technology

Yuki Arase
arase@ist.osaka-u.ac.jp
Osaka University

Takaaki Tsunoda
tsunoda@mibel.cs.tsukuba.ac.jp
University of Tsukuba

Mikio Yamamoto
myama@cs.tsukuba.ac.jp
University of Tsukuba

## ABSTRACT
The frequency of a web search keyword generally reflects the degree of public interest in a particular subject matter. Search logs are therefore useful resources for trend analysis. However, access to search logs is typically restricted to search engine providers. In this paper, we investigate whether search frequency can be estimated from a different resource such as Wikipedia page views of open data. We found frequently searched keywords to have remarkably high correlations with Wikipedia page views. This suggests that Wikipedia page views can be an effective tool for determining popular global web search trends.

## 1. INTRODUCTION
When have the Japanese people shown most interest in the American actress and producer Anne Hathaway recently? To answer this question, we must know the frequency of "Anne Hathaway" being input as a keyword in a web search engine. This frequency peaked on December 12, 2014, following the television broadcast of a movie in which Anne Hathaway played a major role. This fact indicates that people use web search engines frequently to check the details of topics of their interest. Thus, the frequency of a web search keyword generally reflects the prevailing degree of public interest in a specific topic. Studies utilize the frequency of web search keywords to determine trends in public interest. Choi and Varian, for example, demonstrate the use of search engine data to forecast the near-term values of economic indicators [1]. Radinsky et al. present a method for using patterns of web search keywords to predict future events [2]. However, access to search logs is typically restricted to search engine providers. Although some search engines provide search logs via online services such as Google Trends[1], the availability of data from these search engines is fairly limited. For example, we cannot obtain a set of all trend keywords for a specific date. To do so, we would have to burden Google servers by querying all possible keywords that may have been popular on that day. As such, there is a need for a source of open data that can simulate search logs.

[1] https://www.google.com/trends/

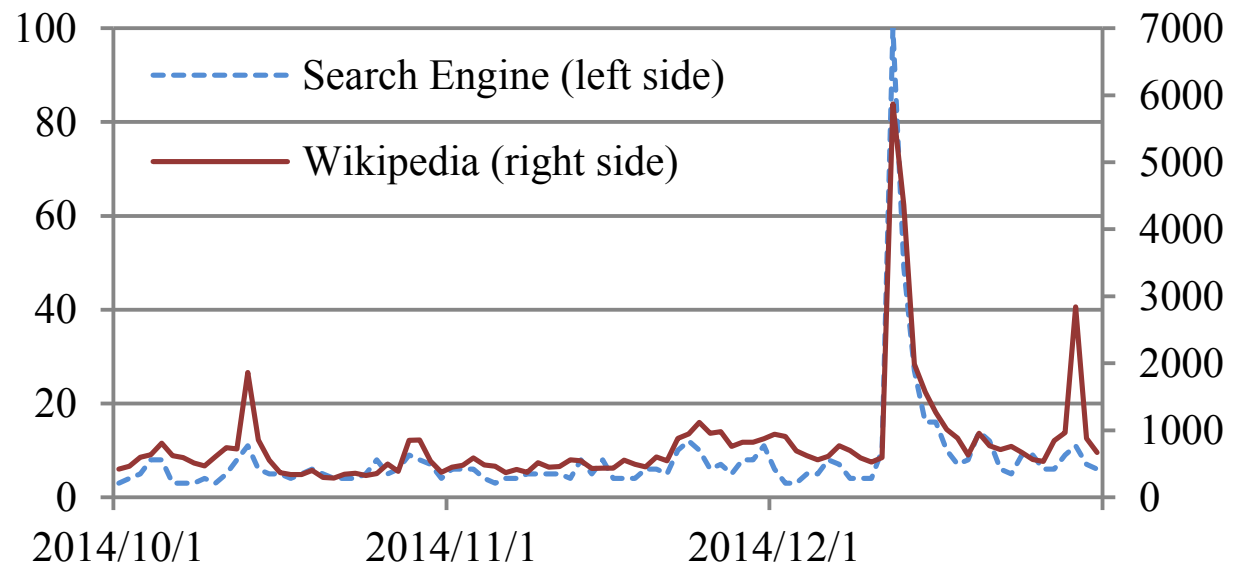

**Figure 1: Daily trend of the keyword "Anne Hathaway" in a search engine (*i.e.*, Google Trends) and Wikipedia.**

We investigate whether search frequency can be estimated from a different resource of open data. We focus on a tendency of Wikipedia web pages appearing in the high rank by many search result sets. Previous studies have reported that users tend to visit more highly ranked search result pages (*e.g.*, [3]). In addition, a recent paper has shown the effectiveness of Wikipedia page views for understanding trends. Specifically, it investigates correlations between BitCoin price and Google Trends or Wikipedia [4]. Based on these studies, we hypothesize that the frequency of accessing Wikipedia (*i.e.*, page views) will reflect search frequency. Figure 1 depicts an example, in which we see remarkably high correlations between the search frequencies of Google Trends and Wikipedia page views. We attempt to test this hypothesis by determining correlations between search frequencies and Wikipedia page views.

## 2. METHODOLOGY AND DATA

### 2.1 Methodology
We calculate the similarities between search frequencies and Wikipedia page views over daily and monthly time spans, because the appropriate span for estimating search frequencies varies according to their applications.

We use the Pearson product-moment correlation coefficient as our measurement tool. In addition, we apply an up/down concordance rate to explicitly consider the increase/decrease in frequencies. In contrast to the Pearson product-moment correlation coefficient, the concordance rate is sensitive to frequency trends. It inputs two sets of time-series data $\mathbf{X} = \{x_1, x_2, \dots, x_n\}$ and $\mathbf{Y} = \{y_1, y_2, \dots, y_n\}$ and computes the following value.

$$UDCR(\mathbf{X}, \mathbf{Y}) = \frac{|\{t \in \{2,3,\dots,n\} | x'_t = y'_t\}|}{n-1},$$

where $x'_t$ is

$$x'_t = \begin{cases} 1 & (x_t - x_{t-1} > 0) \\ 0 & (x_t - x_{t-1} = 0), \\ -1 & (x_t - x_{t-1} < 0) \end{cases}$$

and $y'_t$ is calculated in the same manner.

## 2.2 Data

Since access to search logs is restricted, we must rely on Google Trends to obtain search frequencies. These trends are expressed as the percentage integers of a maximum value for a particular search interval, *i.e.*, the absolute frequency values cannot be known. The trends of low frequency keywords consist mostly of zeros or are not even provided by Google Trends. In our experiment, we collect daily and monthly search frequencies by inputting keywords into Google Trends.

The Wikimedia Foundation, Inc. distributes page view statistics for Wikimedia projects[2]. Each statistic comprises the page path (*e.g.*, "wiki/Anne_Hathaway") and the number of hourly page views. To generate our experimental data, we first prepare a keyword list from which we estimate trends. Second, we correlate each keyword (*e.g.*, "Anne Hathaway") to a page path if the keyword exists as a Wikipedia article title. Third, we calculate the daily and monthly page views from hourly data to obtain the daily and monthly page views for each specific keyword. Finally, we compare those daily and monthly page views with search logs.

For this experiment, we used article titles from the Japanese Wikipedia as keywords to compare their trends in search engines and Wikipedia page views. Specifically, we selected 10,000 keywords that were accessed most often from 2008 to 2014 in the personal name category and obtained their daily and monthly search frequencies and page views. In total, we collected daily frequencies of 2,039 keywords from October to December 2014 and monthly frequencies of 9,757 keywords from 2008 to 2014. These numbers of keywords differ because Google Trends does not provide search frequency statistics for low frequency keywords. Figure 2 shows the distribution of the number of collected keywords in Google Trends. This dataset is published[3].

Our experiment also included cartoon, comic, and movie title keywords. In any case, the analysis results of these keywords were similar to those of the person names reported in this paper.

# 3. RESULTS

As an example, Figure 1 shows the daily trend of search frequency and Wikipedia page views for the keyword "Anne Hathaway". The correlation coefficient between the search frequency and the Wikipedia page views is 0.92, and the concordance rate is 0.54.

Figure 3 and Figure 4 show the average correlation coefficients and concordance rates, respectively. The *x*-axis indicates the access rank in Wikipedia, and the *y*-axis shows the average correlation coefficients or concordance rates. For example, Figure 3 reveals that on a monthly basis, the average correlation coefficient for keywords ranked from 1 to 1,000 in Wikipedia page views is 0.72 and that of keywords ranked from 1,001 to 2,000 is 0.74. This implies that frequently accessed keywords tend to have high correlation coefficients and concordance rates. With respect to the actual number of page views rather than rank, the average correlation coefficient of keywords with more than 1,000 accesses to Wikipedia per day (ranking 838 or higher) are 0.57 daily and 0.72 monthly. The average concordance rates are 0.52 and 0.70, respectively. These results confirm that Wikipedia page views are effective resources for estimating web search trends. On the other hand, the measures for infrequently accessed keywords tend to be low. The reason for this is that Google Trends data consist mostly of zeros. We would like to investigate actual search frequency data in future studies.

[2] http://dumps.wikimedia.org/other/pagecounts-raw/

[3] http://dx.doi.org/10.5281/zenodo.14539

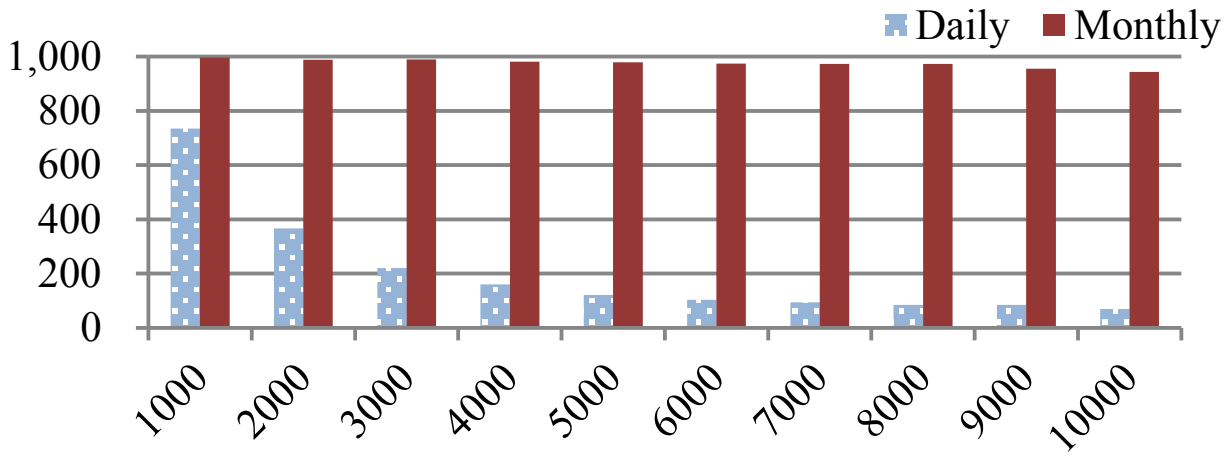

**Figure 2: Distribution of the number of collected keywords in Google Trends: The *x*-axis indicates the access rank in Wikipedia, while the *y*-axis shows the number of keywords whose frequency was obtained by Google Trends.**

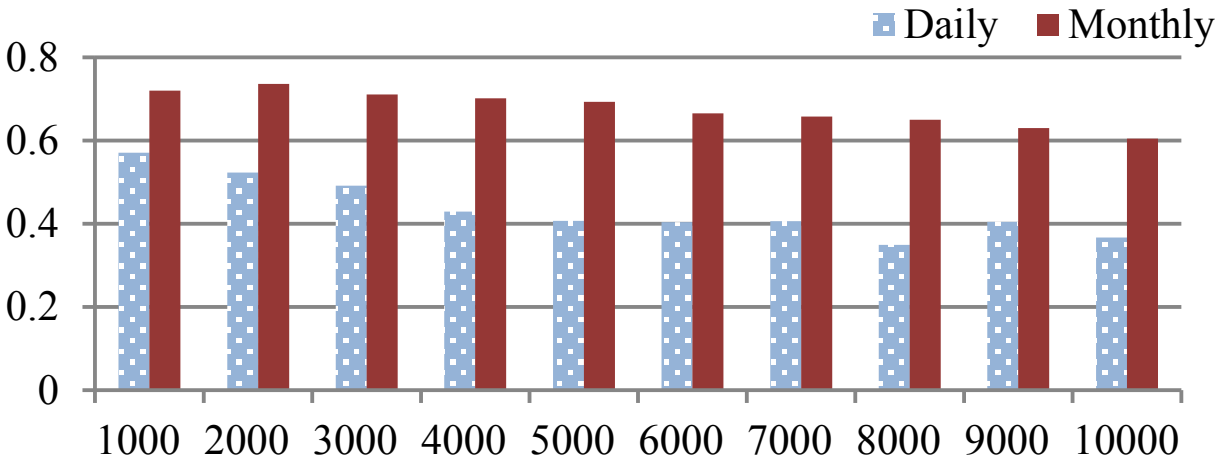

**Figure 3: Average correlation coefficients per access rank in Wikipedia page view.**

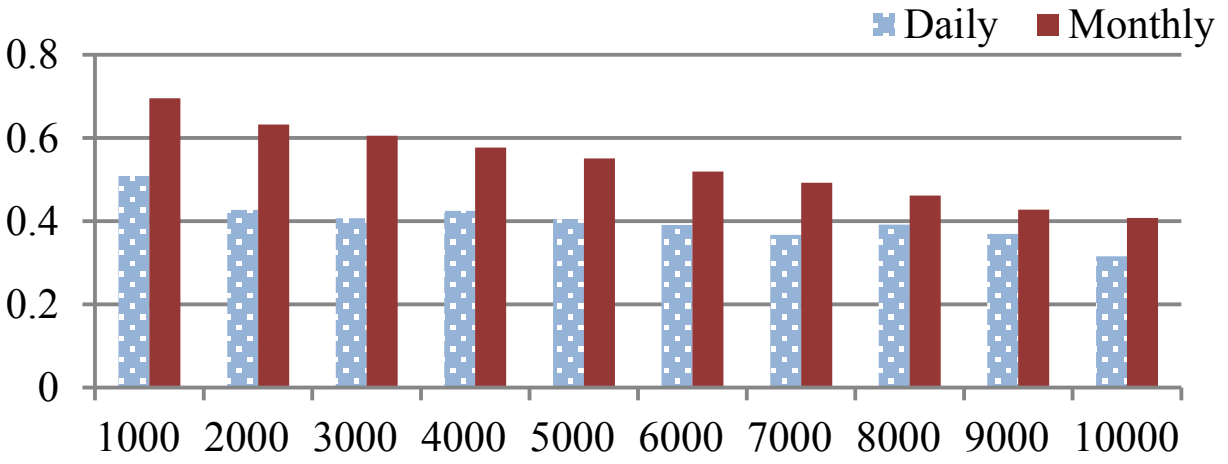

**Figure 4: Average concordance rates per access rank in Wikipedia page view.**

# 4. CONCLUSION

In this paper, we conducted experiments to investigate whether search frequency (*i.e.*, Google Trends) can be estimated from Wikipedia page views of open data. We found remarkably high correlations between them for frequently searched keywords. Specifically, keywords that were accessed more than 1,000 times per day on their corresponding Wikipedia web pages showed a monthly correlation coefficient of 0.72 with the search frequencies obtained by Google Trends. These findings suggest that, at least for high frequency keywords, Wikipedia page views can be effective resources for simulating web search trends.